\documentclass{appolb}

\usepackage{amsmath}
\usepackage{amssymb,amsthm,graphicx,citesort}
\usepackage[ansinew]{inputenc}
\usepackage{stmaryrd}
\usepackage{multibox}
\usepackage{graphicx, epic, eepic, color}
\usepackage{dcolumn}
\usepackage{bm}
\usepackage[ansinew]{inputenc}
\usepackage{times}

\begin{document}

\title{Comments on\\ ``Spin Connection Resonance in Gravitational General
Relativity''}

\author{Gerhard W.~Bruhn\footnote{Department of Mathematics, Darmstadt
    University of Technology, Schlossgartenstr.\ 7, 64289 Darmstadt,
    Germany; email: bruhn@mathematik.tu-darmstadt.de}\\
  Friedrich W.~Hehl\footnote{Institute for Theoretical Physics,
    University of Cologne, 50923 K\"oln, Germany, and Department of
    Physics and Astronomy, University of Missouri-Columbia, Columbia,
    MO 65211, USA; email: hehl@thp.uni-koeln.de}\\
  Arkadiusz Jadczyk\footnote{Center CAIROS, Laboratoire de
    Math\'ematiques Emile Picard, Institut de Math\'ematiques de
    Toulouse, Universit\'e Paul Sabatier, 31062 TOULOUSE CEDEX  9, France; email:
    arkadiusz.jadczyk@cict.fr}}

\date{29 July 2007, {\it file Evans-Resonance06.tex}}

\maketitle

\begin{abstract}
  We comment on the recent article of Evans in this journal
  \cite{Evans1}.  We point out that the equations underlying Evans'
  theory are highly problematic.  Moreover, we demonstrate that the
  so-called ``spin connection resonance'', predicted by Evans, cannot
  be derived {}from the equation he used. We provide an exact solution
  of Evans' corresponding equation and show that is has definitely
  {\em no\/} resonance solutions.
\end{abstract}

\begin{footnotesize}
PACS numbers: 03.50.Kk; 04.20.Jb; 04.50.+h

Keywords: Electrodynamics, gravitation, Einstein-Cartan theory, Evans'
unified field theory
\end{footnotesize}

\section{Introduction}

Over the last years, Evans' papers deal mainly with his so-called
Einstein-Cartan-Evans (ECE) theory, which exists also under the former
name ``Generally covariant unified field theory'' \cite{Evans2}. Evans
aims at a fundamental unified field theory for physics. However, a
long list of serious errors in his theory is well-known, see
\cite{Bruhn1,Bruhn2,Rodrigues,Jadczyk,HehlObukhov}. Evans never tried
to take care of these errors and to improve his theory
correspondingly. In fact, he believes that his theory is flawless.

In our opinion it is clear that Evans' theory has been disproved
already and is untenable, both from a physical and a mathematical
point of view. Nevertheless, he continues to publish papers and
to predict new physical effects. In \cite{Evans1}, Evans foresees a new
``spin connection resonance'' (SCR) effect. The aim of our article is
to take a critical view on \cite{Evans1}.

In Sec.2 we go through Evans' article \cite{Evans1} and point out
numerous mistakes and inconsistencies in the set-up of his theory.
Most of it is known from the literature
\cite{Bruhn1,Bruhn2,Rodrigues,Jadczyk,HehlObukhov}. In Sec.3 we turn
to the new SCR effect, which Evans derives from a certain ordinary
differential equation of second order. Even though the derivation of
this equation is dubious, we start from exactly the same equation as
Evans did and prove that this equation has no resonance type solutions
as Evans claims. This shows that Evans' SCR effect is a hoax.

\section{General comments on Evans' paper}

The paper \cite{Evans1} deals with what the author calls `Cartan
geometry'. The term is not defined in the paper, so the reader has to
guess what the exact meaning is of this term. {}From the content of the
paper it seems plausible that the term means: linear connection in the
tangent bundle of a four--dimension manifold, compatible with the
metric of Minkowskian signature, see also \cite{HehlObukhov} for a
discussion of this `Riemann-Cartan geometry'. The connection may admit
torsion, and the method used is that of Cartan's moving frame (also
known as tetrad or vierbein). In what follows we will assume this
interpretation of the term `Cartan geometry' in our paper. We will
refer to the equations in Ref.\ \cite{Evans1} by using double
parenthesis.

\subsection{Curvature and torsion}

Evans' paper starts with what the author calls ``the second Cartan
structure equation'',
\begin{equation*} \label{eq:sse}
  R^a_{\ b}=D\wedge \omega^a_{\ b}\,,\tag{(1)}
\end{equation*}
and with the second Bianchi identity,
\begin{equation*}\label{eq:sbi}
  D\wedge R^a_{\ b}:=0\,.\tag{(2)}
\end{equation*}
The symbol $D\wedge$ stands, in Evans' notation, for the exterior
covariant derivative, $\omega$ and $R$ are the connection and the
curvature forms, respectively. Eq.(\ref{eq:sse}) represents the {\em
  definition\/} of the curvature form. The second structure equation,
which follows immediately {}from (\ref{eq:sse}) and {}from the
definition of $D$, is given as
\begin{equation*}\label{eq:sseb}
  R^a_{\ b}=d\wedge\omega^a_{\ b}+\omega^a_{\
    c}\wedge\omega^c_{\ b}\,.\tag{(5)}
\end{equation*}
The second Bianchi identity follows {}from ((5)) by exterior
differentiation:
\begin{equation*}\label{eq:bb2}
  d\wedge R^a_{\ b}+\omega^a_{\ c}\wedge R^c_{\ b}-R^a_{\ c}\wedge
  \omega^c_{\ b}:=0\,.\tag{(6)}
\end{equation*}
Torsion is
introduced according to
\begin{equation*}\label{eq:ta}
  T^a=d\wedge q^a+\omega^a_{\ b}\wedge
  q^b\tag{(7)}\,,
\end{equation*}
with the tetrad one-forms $q^a,$ which we interpret, according to the
context, as a local {\em orthonormal} coframe.

\subsection{Objections to the `derivation' of Eqs. ((11)) and ((13))}

Subsequently Evans writes:\medskip

``\dots Eq.((6)) can be rewritten as
\begin{equation*}
  d\wedge R^a_{\ b}=j^a_{\ b}\,,\tag{(10)}
\end{equation*}
\begin{equation*}
  d\wedge{\widetilde R}^a_{\ b}={\widetilde j}^a_{\ b}\,,\tag{(11)}
\end{equation*}
where
\begin{equation*}
  j^a_{\ b}=R^a_{\ c}\wedge\omega^c_{\ b}-\omega^a_{\ c}\wedge
  R^c_{\ b},\tag{(12)}
\end{equation*}
\begin{equation*}
  {\widetilde j}^a_{\ b}={\widetilde R}^a_{\ c}\wedge\omega^c_{\ b}
  -\omega^a_{\ c}\wedge {\widetilde R}^c_{\ b}.\tag{(13)}
\end{equation*}
The tilde denotes the Hodge dual [1--20] of the tensor valued
two--form
\begin{equation*}
R^a_{\ b\mu\nu}=-R^a_{\ b\nu\mu}\,\dots\mbox{''}\tag{(14)}
\end{equation*}

While it is true that ((10)) and ((12)) are a rewriting of ((6)), this
is {\em false\,} for ((11)) and ((13)). Eqs.((11)) and ((13)) do not
follow {}from differential geometry. Especially the combination of
((11)) and ((13)), namely
\begin{equation*}
  d\wedge \widetilde{R}^a_{\ b} = \widetilde{R}^a_{\ c}\wedge\omega^c_{\ b} -
  \omega^a_{\ c}\wedge \widetilde{R}^c_{\ b}\,,
\end{equation*}
cannot be derived {}from the second Bianchi identity ((6)) and does not
hold in general. Indeed, $D\wedge R^a_{\ b}=0$ does {\em not\/} imply
$D\wedge {\widetilde R}^a_{\ b}=0$, since taking the Hodge dual doesn't
commute with $D$.

\subsection{The electromagnetic sector of Evans' theory, the index
  type mismatch}

Eqs.((17)) and ((18)) relate, according to Evans, a generalized
electromagnetic field strength $F^a$ and a potential $A^a$ to the
torsion and the tetrad, respectively,
\begin{equation*}\label{eq:fa}
  F^a=A^{(0)}T^a\,,\tag{(17)}
\end{equation*}
\begin{equation*}\label{eq:aa}
  A^a=A^{(0)}q^a\,,\tag{(18)}
\end{equation*}
where $A^{(0)}$ is, presumably, a universal constant. Evans' next but
one equation is the first Bianchi identity,
\begin{equation*}\label{eq:20}
  d\wedge T^a=R^a_{\ b}\wedge q^b-\omega^a_{\ b}\wedge
  T^b\,.\tag{(20)}
\end{equation*}

Let us look at Evans' motivation for his choices ((17)) and ((18)).
Evans supposed an analogy of $A^a$ and $F^a$ with the Maxwellian
potential one--form $A$ and the field strength two--form $F$ according
to
\begin{equation}\label{subst}
  A \rightarrow A^a\,, \qquad    F\rightarrow F^a\,.
\end{equation}
In Maxwell's theory, $F=d\wedge A$ is then put in analogy to
Cartan's first structure equation (definition of the torsion) $T^a =
D\wedge q^a$.

One serious objection is based on the fact that Evans has not given
any information about the relations between the concrete
electromagnetic fields $F=(\bf E,\bf B)$ in physics and his quadruple
of two--forms $F^0, F^1, F^2,F^3$ and the associated quadruple of
one--forms $ A^0, A^1, A^2, A^3$. Evans himself ignores that problem
of attaching a superscript $a$ to all electromagnetic field quantities
without giving a satisfying explanation of that index surplus.

Evans' attempts to interpret (\ref{subst}) appropriately doesn't
even work in the case of a simple circularly polarized plane (cpp)
wave. His considerations are contradictory and incomplete, and we see
no way to define $ F^0, F^1, F^2,F^3$ and $ A^0, A^1, A^2,$ $A^3$ even
for a bit more complicated field as, e.g., a superposition of
different cpp waves travelling in different directions.  This is not a
mathematical error, but a physical gap, and we doubt that one can find
a general solution of that problem.  Anyway, Evans never presented
such a solution.

Therefore, Evans' analogy $F \leftrightarrow
T^a\,,\mbox{for}\;a=0,1,2,3$, causes a {\em type mismatch} between the
{\em vector} valued torsion two--form $T^a$ and the {\em scalar}
valued electromagnetic field strength two--form $F$.  The analogous
holds for $A \leftrightarrow q^a\,,\mbox{for}\;a=0,1,2,3$ as well.

Evans' whole SCR paper is based on the dubious assumption that
(\ref{subst}), and thus ((17)) and ((18)), make sense in physics.
Without a concrete physical interpretation of (\ref{subst}), Evans'
whole SCR paper is null and void, regardless whether there are other
(mathematical) errors or not.

Moreover, as it was with the second Bianchi identity, so here, Evans'
equations ((23)),((16)), and ((17)), if combined, lead to
\begin{equation}\label{1st-}
  d\wedge {\widetilde T}^a={\widetilde R}^a_{\ b}\wedge q^b-\omega^a_{\
    b}\wedge {\widetilde T}^b\,.
\end{equation}
Eq.(\ref{1st-}), contrary to Evans' statement, is {\em not\/} a
consequence of the first Bianchi identity and does not hold in
Cartan's differential geometry. It represents an additional ad hoc
assumption.

\subsection{The gravitational sector of Evans' theory, objections to
  Eq.((30))}

Eq.((29)) is the field equation of Einstein's general relativity theory,
\begin{equation*}\label{eq:efe}
  G^{\mu\nu}=kT^{\mu\nu}\,,\tag{(29)}
\end{equation*}
after which Evans writes:\medskip

``\dots Eq.(29) is well known, but much less transparent than the
equivalent Cartan equation
\begin{equation*}\label{eq:eeq}
\begin{array}{rcl}D\wedge\omega^a_{\ b}&=&kD\wedge T^a_{\ b}\nonumber\\
&:=&0\,\dots\mbox{''}
\end{array}\tag{(30)}
\end{equation*}

Eq.((30)) is certainly {\em not\/} equivalent to ((29)), and it cannot
be a part of general relativity theory, be it tensorial or in
Cartan form. The reason is very simple: $T^a_{\ b}$ in ((30)) has to
be a one--form. Therefore it should be integrated over a world--line
and not over a hypersurface of four--dimensional spacetime, as it is
done with the energy--momentum tensor. In other words, Eq.((30)) is
simply incorrect since the energy--momentum in exterior calculus is a
covector--valued three--form (or, if its Hodge dual is taken, a {\em
  covector--valued} one--form).

\subsection{The wrong `curvature vector' and the dubious potential equation}

Now Evans turns to the combined equation ((5)) and ((10)),
\begin{equation*}\label{eq:ecb}
d\wedge (d\wedge\omega^a_{\ b}+\omega^a_{\ c}\wedge\omega^c_{\ b})=j^a_{\
b}\,,\tag{(31)}
\end{equation*}
with his comment that {\it in vector notation\/} it gives, in
particular,
\begin{equation*}\label{eq:32}
  \mathbf{\nabla}\cdot{\mathbf
    R}\mathrm{(orbital)}=\mathbf{J}^0,\tag{(32)}
\end{equation*}
with
\begin{equation*}\label{eq:orb}
  \mathbf{R}\mathrm{(orbital)}=R^{0\phantom{1}01}_{\ 1}\mathbf{i}
  +R^{0\phantom{2}02}_{\ 2}\mathbf{j}
  +R^{0\phantom{3}01}_{\ 3}\mathbf{k}\,.\tag{(33)}
\end{equation*}
It is evident that ((32)) is not equivalent to ((31)), if only for the
simple reason that ((31)) involves a three--form, where all indices
must be different {}from each other, while ((32)), with the divergence
operator, involves summation over repeated indices. In ((37)), Evans
evidently attempts to calculate the $(0i)$ component of the curvature
form:
\begin{equation*}\label{eq:37}
\bm{R}^a_{\ b}=-\frac{1}{c}\frac{\partial\bm{\omega}^a_{\ b}}{\partial
t}-\bm{\nabla}\omega^{0a}_{\phantom{0a}\ b}-\omega^{0a}_{\ \
c}\bm{\omega}^c_{\ b}+\bm{\omega}^a_{\ c}\omega^{0c}_{\phantom{0a}\
b}\,.\tag{(37)}
\end{equation*}
This is again incorrect. In fact, starting {}from ((5)), the calculation
of the components $(R_{0i})^a_{\ b},\, \mbox{for}\;i=1,2,3$, yields
\begin{equation}
  (R_{0i})^a_{\ b}=\partial_0(\omega_i)^a_{\ b}-\partial_i(\omega_0)^a_{\
    b}+(\omega_0)^a_{\ c}(\omega_i)^c_{\ b}-(\omega_i)^a_{\
    c}(\omega_0)^c_{\ b}\,.
\end{equation}
Raising the index $0$ of $\omega_0$ in the term
$\partial_i(\omega_0)^a_{\ b},$ as Evans does, is illegitimate,
because the metric component $g^{00}$ of the Schwarzschild metric,
which Evans considers, is {\em not\/} a constant function of the
variables $x^i.$ The sign in front of the time derivative in ((37)) is
also wrong.

Then in ((42)), when restricting to the static case, Evans `forgets'
one of the quadratic terms of his erroneous ((37)):
\begin{equation*}\label{eq:42}
  \bm{R}^a_{\ b}=-\bm{\nabla}\omega^{0a}_{\phantom{0a}\ b}+\bm{\omega}^a_{\
    c}\omega^{0c}_{\phantom{0a}\ b}\,.\tag{(42)}
\end{equation*}
Again, this is wrong, since now $\bm{R}^a_{\ b}$ is not in the Lie
algebra of the Lorentz group. The same error applies to ((44)), where
$\omega^{0a}_{\phantom{0a}\ b}$ is substituted by $\Phi^a_{\ b}$,
\begin{equation*}\label{eq:44}
\bm{R}^a_{\ b}=-\bm{\nabla}\Phi^{a}_{\ b}+\bm{\omega}^a_{\ c}\Phi^c_{\
b}\,.\tag{(44)}
\end{equation*}
Then Evans adds:\medskip

``\dots It is convenient to use a negative sign for the vector part of
the spin connection, so
\begin{equation*}\label{eq:45}\bm{R}^a_{\ b}=-\left(\bm{\nabla}
\Phi^{a}_{\ b}+\bm{\omega}^a_{\ c}\Phi^c_{\
b}\right)\dots\mbox{''}\tag{(45)}
\end{equation*}

This is another evident and grave error. Since the sign of the
connection form is not a question of `convenience' in the theory of
gravity, where the curvature tensor contains both linear and quadratic
terms in the connection. Changing the sign of the connection forms
changes its curvature in an essential way.

Using incomprehensible and sometimes evidently wrong reasonings, such
as skipping one term when going {}from ((37)) to ((42)), as we saw
above, Evans postulates a potential equation ((63)) for an
unidentified variable $\Phi$ for the case of the Schwarzschild
geometry. We shall discuss the ``electromagnetic analogue of
Eq.(63)'', namely Eq. ((65)), in the following section.

\section{The Resonance Catastrophe}

In the lines after ((31)), Evans writes:\medskip

``It is shown in this section that Eq.(31) produces an
infinite number of resonance peaks of infinite amplitude in the gravitational
potential [2--20]. To show this numerically, Eq.(31) is developed in vector
notation\dots''\medskip

This is an unfounded claim followed by {\em no proof} and no numerical
results either. In addition the claim is erroneous as we shall see
below. At the very end of his article Evans at last arrives at the
topic `resonance' that is already announced in the title of his paper.
He reports:\medskip

``\dots The electromagnetic analogue of Eq.(63) is
\begin{equation*}
  \frac{\partial^2 \phi}{\partial r^2} -
  \frac{1}{r} \frac{\partial \phi}{\partial r} +
  \frac{1}{r^2} \phi=- \frac{\rho (0)}{\epsilon_o} \cos(\kappa r) \tag{(65)}
\end{equation*}
which has been solved recently using analytical and numerical methods
[2--20].  These solutions for $\phi$ and $\Phi$ show the presence of
an infinite number of resonance peaks, each of which become infinite
in amplitude at resonance.''\medskip

Evans' efforts (together with H. Eckardt) with respect to the
resonance of ((65)) are available on his website.  He attempts to find
values of the parameter $\kappa$ that yield resonances of the right
hand side of ((65)) with the eigensolutions of this Euler type
ordinary differential equation (ODE). However, the eigensolutions of
the associated homogeneous ODE are well-known. The eigenspace is
spanned by the special solutions
\begin{equation}\label{eigensol}
  \phi_1 = r \qquad\mbox{and}\qquad \phi_2 = r \log r\, .
\end{equation}
Resonance means that the driving term $\cos(\kappa r)$ belongs to the
eigenspace, i.e., is a linear combination, with constant coefficients,
of the functions $\phi_1$ and $\phi_2$ for any value of the parameter
$\kappa$. Obviously this is not the case.

\begin{figure}\label{fig1}
\includegraphics[width=10cm]{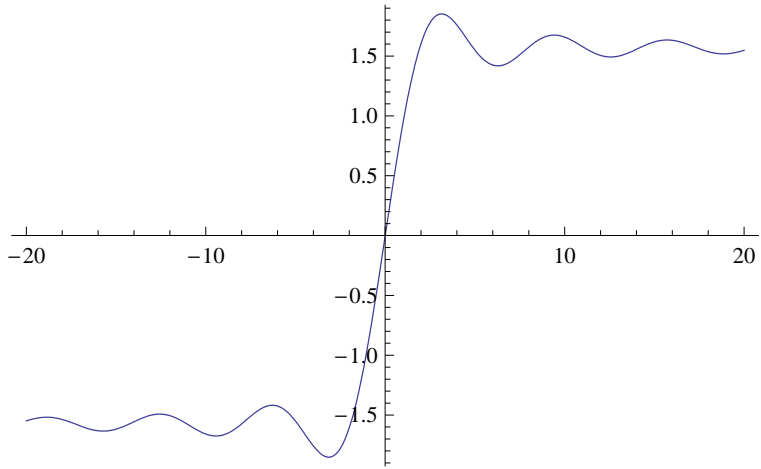}
\caption {Graph of the sine integral ${\rm Si(z)}$.}
\end{figure}

Moreover, the general solution of ((65)) can be calculated. With the
help of Mathematica, we obtain
\begin{equation}\label{sol}
  \phi(r) = c_1 r + c_2 r \log r
  - \frac{\rho (0)}{\epsilon_o} \,\frac{r}{\kappa}\, {\rm Si}(\kappa r)\,,
\end{equation}
where  ${\rm Si}$ denotes the {\em sine integral} function defined by
\begin{equation}
  {\rm Si}(z) := \int_0^z \frac{\sin t}{t} dt
\end{equation}
 \noindent
for real z satisfying the estimate
\begin{equation}
  \left| {\rm Si}(z) \right| \le \min(|z|,2)\,.
\end{equation}
The graph of  ${\rm Si(z)}$ is displayed in Fig.1.

Thus, the $\kappa$ dependent part of the solution (\ref{sol})
satisfies the estimate
\begin{equation}
  \left| \frac{\rho (0)}{\epsilon_o}\, \frac{r}{\kappa}\, {\rm
      Si}(\kappa r) \right|
  \le \frac{\rho (0)}{\epsilon_o}\, \min (r^2, \frac{2r}{\kappa})\,.
\end{equation}
Consequently, the general solution of (65) is bounded for all real
values of $\kappa$ and $r$. For no value of $\kappa$, we will have a
resonance of the right-hand-side of ((65)) with the eigensolutions
(\ref{eigensol}).

However, Evans \& Eckardt apply a lot of their specific `new math': an
inadmissible rotation of the complex plane of eigenvalues by an angle
of $90^\circ$ and multiplication by the imaginary unit $i$, among
other peculiarities, see \cite{Bruhn2} for details. Evans \& Eckardt
succeed in detecting resonance peaks, unattainable to all who are
using standard mathematics only.  \medskip

{\sl There are no resonance peaks at all, quite apart from the errors
  in Evans' theory previous to his equations ((63)) and ((65)).}

\end{document}